\documentclass[prb,twocolumn,superscriptaddress,floatfix,longbibliography,nofootinbi]{revtex4-2}

\usepackage{mathtools}
\usepackage{latexsym}
\usepackage{graphicx}
\usepackage{float}
\usepackage{dcolumn}
\usepackage{bm}
\usepackage{amssymb}
\usepackage{amsmath}
\usepackage{mathtools}
\usepackage[space]{grffile}
\usepackage{xcolor}
\usepackage{comment}
\usepackage{cancel}
\usepackage[export]{adjustbox}
\usepackage{longtable}

\begin{document}


\title{Realizing supersymmetry in a digitized quantum device}

\author{Sutapa Samanta}
 \affiliation{Department of Physics and Astronomy, Western Washington University, Bellingham, Washington 98225, USA}

\author{Jian-Xin Zhu}
\affiliation{Theoretical Division, Los Alamos National Laboratory, Los Alamos, New Mexico 87545, USA}
\affiliation{Center for Integrated Nanotechnologies, Los Alamos National Laboratory, Los Alamos, New Mexico 87545, USA}
 
\author{Armin Rahmani}
\affiliation{Department of Physics and Astronomy, Western Washington University, Bellingham, Washington 98225, USA}
\affiliation{Advanced Materials Science and Engineering Center, Western Washington University, Bellingham, Washington 98225, USA}

\date{\today}

\begin{abstract}
The tricritical Ising model serves as an example of emergent spacetime supersymmetry, which can arise in condensed matter systems. In this work, we present a variational quantum algorithm to create this tricritical state on a digitized quantum computer. We successfully prepare the state on an IBM quantum device and verify the critical correlations of supercurrents associated with the state. Additionally, we examine how other signatures of supersymmetry, namely universal gap ratios and the central charge, can be probed on quantum devices.

\end{abstract}

\maketitle

\section{Introduction}

Spacetime supersymmetry (SUSY) was proposed decades ago as a fundamental principle of nature~\cite{GERVAIS1971632,VOLKOV1973109}, but its experimental confirmation has remained elusive. Interestingly, certain condensed matter systems can exhibit emergent spacetime SUSY in their low-energy behavior, even though SUSY is absent at the microscopic level for underlying electrons~\cite{Grover2014,Rahmani2015,Rahmani2015a, Zhu2016,Ejima2016,OBrien}. Despite their theoretical appeal, these models have yet to be realized in physical materials. Synthetic platforms, particularly quantum computing devices, have shown great promise in emulating condensed matter systems. These quantum platforms not only offer the potential to demonstrate the properties of complex condensed matter models beyond purely theoretical studies but also enable the simulation of phenomena arising in these models that are challenging to calculate using classical computers or observe in conventional experiments.

This paper focuses on the realization and demonstration of emergent spacetime SUSY on a quantum computing device. Physically demonstrating SUSY on these quantum devices, while currently limited to small system sizes, paves the way for probing questions beyond the reach of classical computational tools in future generations of quantum hardware. For instance, the behavior of highly excited states, the long-time limit of non-equilibrium dynamics, and the effects of long-range perturbations are not easily accessible to computational methods such as the density-matrix renormalization group (DMRG). Thus, our work lays the groundwork for addressing such questions as quantum hardware continues to advance.

The tricritical Ising (TCI) model represents the simplest superconformal field theory~\cite{PhysRevA.4.1071,FRIEDAN198537,KASTOR1989590,ZAMOLODCHIKOV1991524}. Consequently, emergent SUSY can arise in condensed matter systems at a quantum phase transition governed by the TCI field theory for strongly interacting many-body Hamiltonians. Notably, it has been demonstrated that simple one-dimensional models of Majorana fermions can host the TCI transition. The minimal model exhibiting this physics involves nearest-neighbor two-Majorana hybridization and interactions among four consecutive Majorana sites in a one-dimensional chain~\cite{Rahmani2015}. An alternative model proposed by O'Brien and Fendley (OF) includes interactions among four Majoranas within a cluster of five consecutive sites~\cite{OBrien}. The OF model realizes the same TCI physics, but requires a much smaller interaction strength, thereby reducing finite-size effects.

In this work, we introduce a variational quantum circuit designed to generate both the ground and first excited states of the OF model. The same variational circuit structure can be employed in variational quantum optimization targeting both states. Using the IBM quantum hardware, we observe signatures of the tricritical state through the correlation functions of the supercharge operators. While the excited-state variational algorithm can produce states with relatively high fidelity, the error levels remain too large to resolve the subtle universal gap ratios. Additionally, we investigate the relationship between the central charge of this model and the Shannon entropy in different bases. A conjectured relationship~\cite{Alcaraz:2013ara,Alcaraz:2014tfa} holds well, at least approximately, enabling an estimate of the central charge from the measurement of the Shannon entropy, which exhibits remarkably small finite-size effects in the $X$ basis.

The outline of the remainder of this paper is as follows. In Sec.~\ref{sec:model}, we review the OF model, focusing on features that are important for its realization and verification in a quantum device. In Sec.~\ref{sec:algorithm}, we describe the variational quantum algorithm for creating the states. We quantify the quality of state preparation in terms of the wave function overlap and excess energy. In Sec.~\ref{sec:hardware} , we measure the correlation functions of one of supercurrent operators on the IBM quantum device for a state created with our variational algorithm, demonstrating its agreement with the expected critical exponent. Sec.~\ref{sec:cc} discusses Shannon entropies as a means of probing the central charge of the TCI model, applied to our variational quantum state. We conclude the paper in Sec.~\ref{sec:conc}. Two appendices discuss the supercurrent correlators with periodic boundary conditions and the excited-state variational quantum algorithms for probing gap ratios.

\section{O’Brien-Fendley model}\label{sec:model}
\noindent We review the OF model~\cite{OBrien}, which consists of the transverse-filed Ising terms with an additional three spin interactions. The Ising part of the Hamiltonian
\begin{align}
H_I= -\sum_{j=1}^{L} \left(\sigma_j^x +\sigma_j^z\sigma_{j+1}^z\right)\,
\end{align}
has a $\mathbb{Z}_2$ symmetry, i.e., it commutes with the spin-flip operator $\prod_j \sigma_j^x$. 
The three-spin interaction term is given by 
\begin{align}
    H_3 = \sum_{j=1}^{L}\left( \sigma_j^x\sigma_{j+1}^z\sigma_{j+2}^z + \sigma_j^z\sigma_{j+1}^z\sigma_{j+2}^x\right)\ .
\end{align}
The Hamiltonian has a simple form in terms of Majorana fermions.  After a Jordan-Wigner transformation $\gamma_{2j-1} = \sigma_j^z\prod_{k=1}^{j-1} \sigma_k^x$, and $ \gamma_{2j} = -i\sigma_j^x\gamma_{2j-1}$,
we can write $H_I =i \sum_a \gamma_a \gamma_{a+1}$,  and $H_3 = -\sum_a \gamma_{a-2}\gamma_{a-1}\gamma_{a+1}\gamma_{a+2}$. A linear combination of $H_I$ and $H_3$ can be written as a sum of squares of two fermionic Hermitian operators up to a constant energy shift as follows:
\begin{align}
    H = (Q^+)^2 + (Q^-)^2 = 2\lambda_I H_I + \lambda_3 H_3 + E_0,
\end{align}
where $\lambda_3>0$, $E_0 = L(\lambda_I^2 + \lambda_3^2)/\lambda_3$ and
\begin{align}
    Q^\pm = \frac{1}{2\sqrt{\lambda_3}}\sum_a (\pm 1)^a(\lambda_I \gamma_a \pm i\lambda_3 \gamma_{a-1}\gamma_a\gamma_{a+1})\ .
\end{align}
In Ref. \cite{OBrien}, it was shown that this model realizes TCI model when the couplings are tuned to the value $\lambda_3/\lambda_I = 0.856$.
The lattice analogs  of the supersymmetric currents are
\begin{align}\label{supercurrents}
&G_j = \lambda_I (\gamma_{2j-1} + \gamma_{2j}) +i\lambda_3 (\gamma_{2j-2} +\gamma_{2j+1}) \gamma_{2j-1} \gamma_{2j},\nonumber\\
&\psi_j =  \lambda_I (\gamma_{2j-1} - \gamma_{2j}) +i\lambda_3 (\gamma_{2j-2} -\gamma_{2j+1}) \gamma_{2j-1} \gamma_{2j}.
\end{align}
The correlation $C_{jk} = \langle \psi_j\psi_k\rangle$ and $D_{jk} =  \langle G_j G_k\rangle$ obey the scaling relations $A|j-k|^{-1.4}$ and $B|j-k|^{-3}$ respectively. 
\begin{figure}[]
\centering
\includegraphics[width=0.48\textwidth]{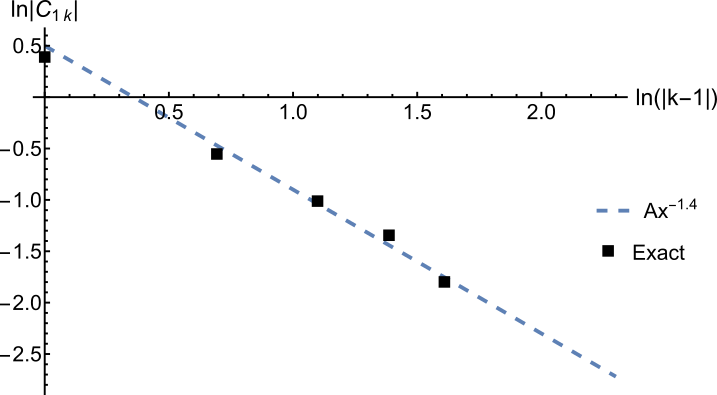}
\caption{Scaling behavior of $C_{jk}$ at the tricritical point for a small 8-qubit system with open boundary condition, obtained by exact diagonalization. The critical exponent emerges in a very small system. }\label{fig:scalingL8}
\end{figure}
The correlation functions discussed above act as a diagnostic for supersymmetry in the system's ground state. A key advantage of the OF model, compared to similar models such as that in Ref. \cite{Rahmani2015}, is that its scaling behavior remains observable even for small system sizes, making it well suited for implementation on near-term noisy quantum devices. In Fig.~\ref{fig:scalingL8}, we show $C_{ij}$ in a system of size $L=8$ with open boundary conditions, obtained via exact diagonalization. The behavior of $C_{ij}$ with periodic boundary conditions is discussed in Appendix~\ref{periodic}.

\section{Variational quantum circuit for state preparation}\label{sec:algorithm}
We employ a variational quantum algorithm to prepare the ground state of the model at the TCI point. The algorithm is based on a variational circuit ansatz, designed with insights from the system's Hamiltonian, as discussed below.

If we wanted to reach the ground state via an adiabatic state transformation, after Trotterizing the evolution, we would apply successive layers of evolution, each corresponding to a Trotter step and incorporating all terms in the Hamiltonian. The single- and two-qubit terms in $H_I$ map directly onto simple quantum gates. However, decomposing the three-qubit interactions in $H_3$ would substantially increase the depth of the circuit. To address this, we retain the terms in $H_I$ but replace the three-qubit interactions with simple single-qubit $Z$ rotations. Unlike $X$ rotations that appear already from $H_I$, a field in the $Z$ direction is missing in the Hamiltonian, and $Z$ rotations are expected to expand the reachable set of the variational ansatz. Instead of adiabatic evolution, we then promote all rotation angles to variational parameters, in line with the approach used in variational eigensolvers and quantum approximate optimization algorithms~\cite{Peruzzo2014,PhysRevA.92.042303,McClean_2016,farhi2014,PhysRevX.7.021027}, using fewer layers than adiabatic evolution.
\begin{figure}[ht]
\centering
\includegraphics[width=0.49\textwidth]{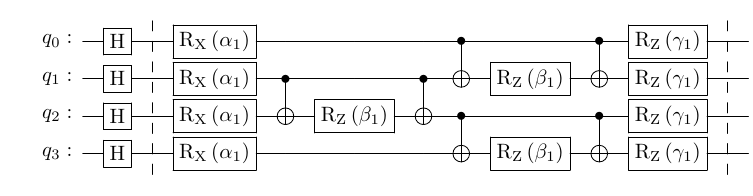}
\caption{Single layer ansatz circuit for a 4-qubit system}\label{circuit}
\end{figure}

A single layer of the ansatz circuit for a 4-qubit system is shown between the dashed barriers in Fig.~\ref{circuit}, which represents the unitary operator $U(\alpha,\beta,\gamma)=  U_{\gamma} U_{\beta} U_{\alpha}$ with $U_{\alpha} = e^{ -i\alpha\sum_j \sigma_j^x }$, $U_{\beta}  =  e^{ -i\beta\sum_j \sigma_j^z\sigma_{j+1}^z }$, and $U_{\gamma} =  e^{ -i\gamma\sum_j \sigma_j^z }$. The Hadamard gates at the beginning create an initial state $|\psi_{\rm in}\rangle = |\phi\rangle \otimes |\phi\rangle\otimes \cdots |\phi\rangle$, where $|\phi\rangle=\left(|0\rangle+|1\rangle\right)/\sqrt{2}$ when all qubits are initialized to $|0\rangle$. Note that for the first layer, the $R_x$ gate does not change the state, as the initial state is their eigenvector. They do become important, however, for the next layers. 

If we use $M$ layers, we obtain a variational wave function
\begin{equation}
  |\psi_{\rm var}\rangle=\prod_{m=1}^M U(\alpha_m,\beta_m,\gamma_m)|\psi_{\rm in}\rangle.  \end{equation}
Variational quantum algorithms targeting the ground state minimize the energy—specifically, the expectation value of the target Hamiltonian—using a hybrid approach in which the expectation value is measured on a quantum device. Here, we simulate that process (the system sizes are small enough that the expectation value can be calculated for the variational state). We also study the direct maximization of the wavefunction overlap with the target state directly. Thus, we have two sets of cost functions based on `wave function overlap maximization' and `energy minimization.' These optimizations provide the values of the variational parameters $\alpha_m, \beta_m$, and $\gamma_m$ (note that $\alpha_1$ is absent as mentioned above). For wave function optimization, we seek to obtain $~99\%$ with the ground state.

As expected, adding more layers strengthens the ansatz by expanding its reach within the Hilbert space. However, more layers also increase the complexity of classical optimization and the susceptibility of quantum evolution to noise and decoherence. We have found that the number of layers needed to obtain a $~99\%$ overlap scales linearly with the system size, i.e., $M=L$ for periodic boundary conditions. For the system with open boundary conditions, we need $M=L+1$ layers to generate the same overlap. 

\begin{figure}[ht]
\centering
\vspace{4mm}
\includegraphics[width=0.4\textwidth]{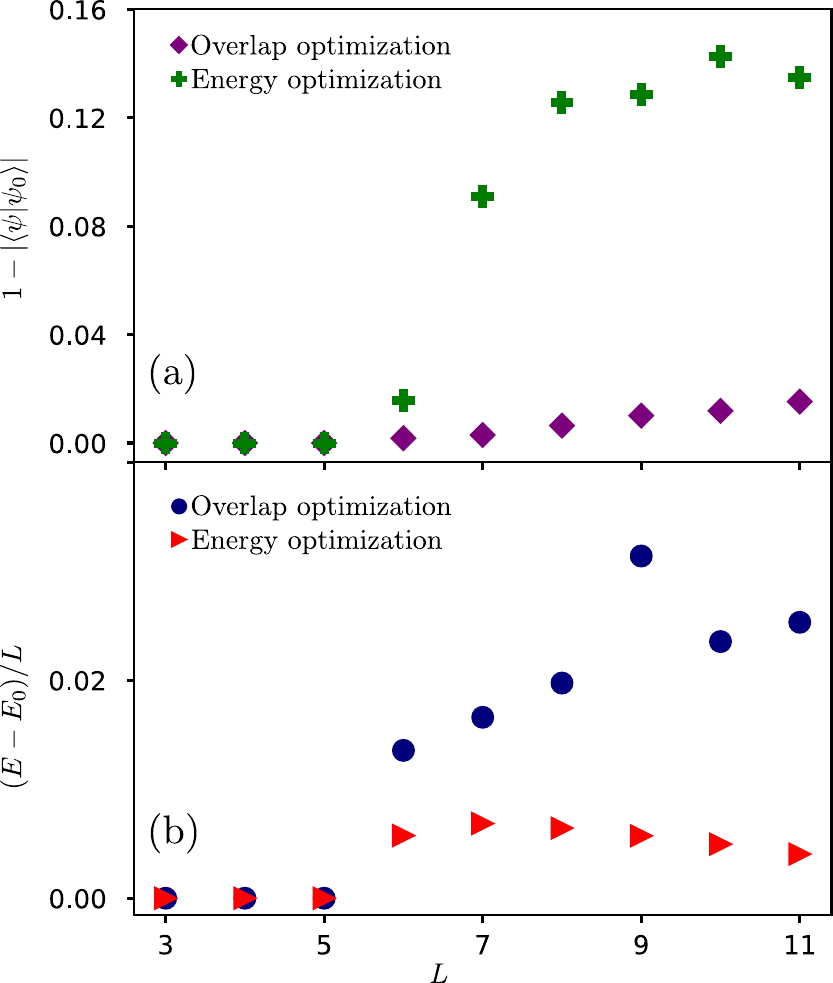}
\caption{Quality of ground state generation using the ansatz. Panel (a) depicts the deviation from perfect groundstate overlap as a function of system size, while panel (b) shows the excess energy density of the generated state. Data from both overlap maximization and energy minimization are presented.}\label{Groundstate}
\end{figure}
In Fig.~\ref{Groundstate}, we assess the effectiveness of the optimized ansatz circuit in approximating the ground state. Let $|\psi_0\rangle$ be the exact ground state of the system and $|\psi\rangle$ the state generated by the variational algorithm. The quantity $1-|\langle\psi|\psi_0\rangle|$ then represents the deviation from the exact ground state. Similarly, we consider the excess energy density $(E-E_0)/L$, where $E_0$ is the exact energy of the ground state, $E$ is the energy of the state generated from optimization, and $L$ is the size of the system.

As shown in Fig.~\ref{Groundstate}(a), the overlap is above $99\%$ when we perform the overlap optimization. Energy optimization cannot reach such high overlaps for larger systems due to convergence issues and gives an overlap of around $85\%$ for the largest systems. However, most of the error comes from low-energy states, and indeed, we obtain lower excess energy. In Fig.~\ref{Groundstate}(b), we show the excess energy density. In contrast to panel (a), energy optimization produces a better result here than overlap optimization. We note that the results of both optimization methods coincide for smaller system sizes.

\section{Quantum hardware implementation and measurement of supercurrent correlators}
\label{sec:hardware}

As a signature of the supersymmetric state, we measure the expectation values of the supercurrent operators on the quantum device. 
Because the qubits in a quantum device are not inherently fermionic, observing the scaling behavior on a quantum computer requires expressing the correlation functions in terms of spin operators via the Jordan-Wigner transformation. Under this mapping, the operators in~\eqref{supercurrents} become sums of Pauli strings. For instance, assuming $\lambda_I=1$, the explicit form of $C_{1k}$ for $3<k<L$ is given by
\begin{widetext}
\begin{align}
 &C_{1k} = i \sigma_1^y \prod_{m=2}^{k-1} \sigma_m^x \sigma_k^z - i\sigma_1^z\prod_{m=2}^{k-1} \sigma_m^x\sigma_k^z + i \sigma_1^y \prod_{m=2}^{k-1} \sigma_m^x \sigma_k^y - i\sigma_1^z\prod_{m=2}^{k-1}\sigma_m^x \sigma_k^y +i\lambda_3^2 \left( \sigma_1^x\sigma_2^y\prod_{m=3}^{k-2}\sigma_m^x\sigma_{k-1}^y\sigma_k^x + \sigma_1^x\sigma_2^y\prod_{m=3}^{k-1} \sigma_m^x\sigma_{k+1}^z\right)\nonumber\\
 &+i\lambda_3 \left(\sigma_1^y\prod_{m=2}^{k-2} \sigma_m^x \sigma_{k-1}^y \sigma_k^x - \sigma_1^z\prod_{m=2}^{k-2}\sigma_m^x \sigma_{k-1}^y \sigma_k^x + \sigma_1^y \prod_{m=2}^{k-1} \sigma_m^x \sigma_{k+1}^z - \sigma_1^z\prod_{m=2}^{k-1} \sigma_m^x\sigma_{k+1}^z + \sigma_1^x\sigma_2^y\prod_{m=3}^{k-1}\sigma_m^x \sigma_k^z +\sigma_1^x\sigma_2^y\prod_{m=3}^{k-1}\sigma_m^x \sigma_k^y\right)\ .
\end{align}
\end{widetext}

Using an ansatz circuit with optimized variational parameters, we prepare the ground state on quantum hardware. The correlators, such as $C_{jk}$, can then be expressed as linear combinations of Pauli strings. Each Pauli string requires a separate measurement. Since IBM quantum devices make measurements on the $z$-basis  by default, additional single-qubit rotation gates must be applied at the end of the circuit when measuring other Pauli operators in the string.

\begin{figure}[ht]
\centering
\includegraphics[width=0.45\textwidth]{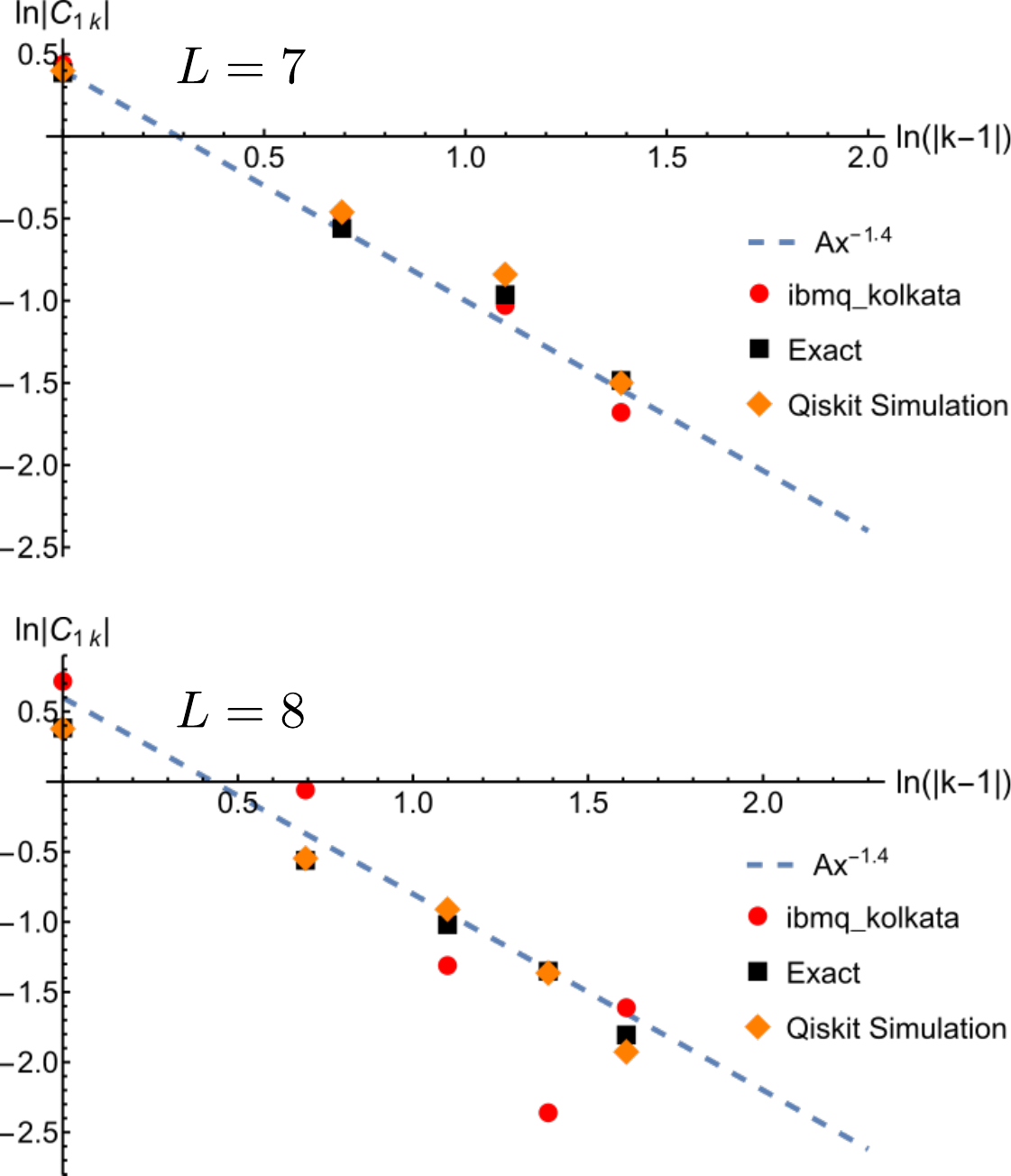}
\caption{Plot of the correlation function $C_{jk}$ as a function of $|j-k|$ in ibmq\_kolkata device for two system sizes.}\label{scalingibmkolkata}
\end{figure}

Using the values of the variational parameters from wavefunction optimization for an open chain and employing complex error mitigation, we reproduce the scaling behavior of the correlation function $C_{jk}$ in ibmq\_kolkata device. The results are shown in Fig.~\ref{scalingibmkolkata}. Qiskit simulation of the variational quantum circuit produces results that agree with the expected critical behavior. For $L=7$, the results for quantum hardware are in excellent agreement with the simulations. The hardware results for $L=8$ are affected more severely by noise.

\section{Probing the central charge on quantum computers}\label{sec:cc}
In addition to the critical exponents of the correlation functions, the central charge $c=7/10$ of the TCI point is a key signature of this state and can serve to verify SUSY. In conformal field theory, the central charge is closely related to the entanglement entropy~\cite{Calabrese_2004}. Thus, a numerical calculation of the entanglement entropy is often used to extract the central charge. Unfortunately, measurement of the entanglement entropy using full-state tomography on a quantum device is challenging~\cite{Cramer2010}. Ideas based on shadow tomography have proved promising for measuring entanglement entropy~\cite{Huang2020,PhysRevA.99.052323,PhysRevA.99.052323,Aaronson2020,PRXQuantum.2.030348,PhysRevLett.133.130803}. Such computation still needs substantial quantum resources due to the large number of measurement.

However, a conjectured {\it approximate} relationship has been proposed for a 1+1$D$ conformal field theory between the basis-dependent Shannon entropy and the central charge, which can be used as an approximate proxy to probe the central charge without the need for many measurements~\cite{Alcaraz:2013ara,Alcaraz:2014tfa,Stephan:2014nda}.

We employ the computation of Shannon mutual information to extract the central charge of the model. The Shannon entropy  of a system is given by
\begin{align}\label{shannon_entropy}
S =-\sum_\sigma p_\sigma \ln p_\sigma ,
\end{align}
where $p_\sigma =|\langle g|\sigma\rangle|^2$ with $|g\rangle$ being the ground state of the system and $|\sigma\rangle$ is basis state of the Hilbert space. The sum runs over all possible configurations of this chosen basis. The leading term obeys the volume law, i.e., for a one-dimensional chain, it is proportional to the system length. The subleading term, however, is approximately related to the central charge. To extract this central charge, we divide the system into two parts, $A$ and $B$, with lengths $\ell$ and $(L-\ell)$, respectively. We then determine the Shannon mutual information.
\begin{align}\label{mutual_information}
    I(A,B) = S(A) +S(B) - S(A\cup B)\ ,
\end{align}
where $S(A)$ is the Shannon entropy of the subsystem $S(A) = -\sum_\mu p_\mu \ln p_\mu $, with 
$p_\mu =\langle\mu|\rho_A|\mu\rangle,\ $
where $\rho_A = \text{Tr}_B |g\rangle\langle g|$ is the reduced density matrix for the subsystem $A$. The mutual information $I(A,B)$ in \eqref{mutual_information} has no linear dependence on $\ell$. It was proposed in~\cite{Alcaraz:2013ara} that Shannon mutual information for a system with periodic boundary conditions is described by the following expression:
\begin{align}
    I(\ell,L) = \frac{c}{4} \ln\left(\frac{L}{\pi}\sin\left(\frac{\pi \ell}{L}\right)\right) + B,
\end{align}
where $c$ is the central charge of the underlying CFT and $B$ is a nonuniversal constant. Although later studies, careful numerical checks on the transverse field Ising model, which maps to free fermions, showed that the relationship is not exact but a good approximation~\cite{Stephan:2014nda}, it is useful to check whether the approximate relationship holds in a strongly interacting model like the TCI model.

\begin{figure}[t]
\centering
\includegraphics[width=0.48\textwidth]{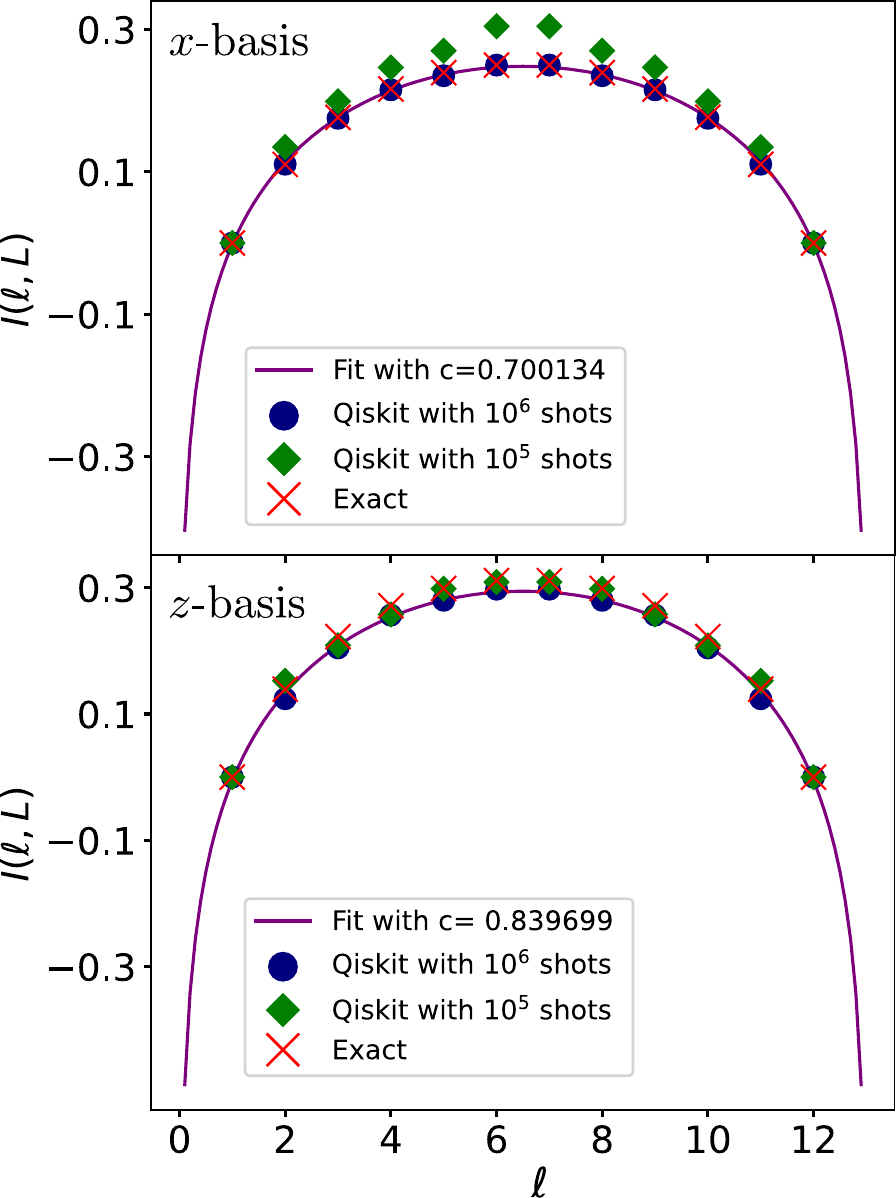}
\caption{Shannon mutual information $I(\ell,L)$ as a function of the sub-system size $\ell$ for a periodic system of size $L=13$. The solid line is the fit with the sub-leading term of Shannon entropy. For Shannon entropies in the $x$ basis (top panel),  we obtain a value of central charge equal to $0.700134$ from the fit. The $z$-basis entropies, however, yield $c=0.8397$ for this system size. }\label{centralchargeL13}
\end{figure}

Here, we explore the relationship between the central charge and the Shannon mutual information, an easy-to-measure quantity on quantum computers, for the O'Brien - Fendley model, both using ground states obtained from exact diagonalization and generated by the variational algorithm. We first start with a system size $L=13$ and consider both the numerically exact ground state and the state obtained from our variational circuit using the qiskit simulator. The Shannon mutual information in the $x$-basis gives a central charge very close to the theoretical value of $c=7/10$. We found, however, that results from the variational quantum circuit need a relatively large number, i.e., $10^6$ measurement shots to converge. The fit for Shannon entropies in the $z$-basis, however, is not in agreement with $c=7/10$ for this system size. It appears that the approximate relationship with the central charge either breaks down for the $z$-basis results or is subject to stronger finite-size effects. As we show below, the latter is the case.

\begin{figure}[!]
\centering
\includegraphics[width=0.48\textwidth]{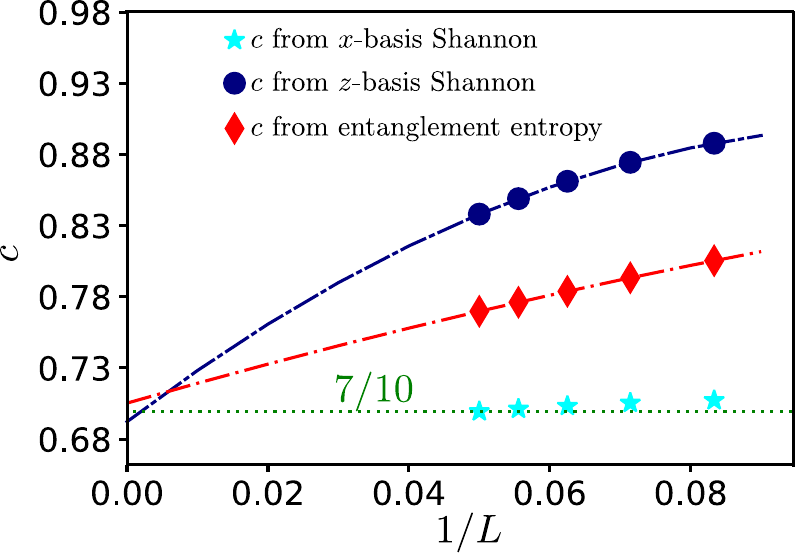}
\caption{Central charge obtained form Shannon entropy calculation in $z$-basis can be fitted with the quadratic function $0.692+ 3.77 (1/L) - 17.05  (1/L)^2$, where $L$ is the system size. The central charge from entanglement entropy fits $0.705+ 1.42 (1/L) - 2.68 (1/L)^2$.
Both extrapolations to infinite system size lead to values close to the theoretical value $c=7/10$.}\label{centralcharge}
\end{figure}

In Fig.~\ref{centralcharge}, we compare the central charge extracted from the $z$-basis Shannon entropy with $c$ obtained from the entanglement entropy, which is known to converge to $7/10$ in the thermodynamic limit. Although the results from $z$-basis Shannon entropy exhibit large finite-size effects, we observe similar effects in the entanglement entropy. A quadratic fit to $1/L$ leads to similar results consistent with a $c=7/10$ central charge. While Shannon entropy in both $x$ and $z$ basis are governed by the same approximate relationship with the central charge, the $x$-basis result has weaker finite-size effects and better serves as verification of SUSY in system sizes accessible to current hardware.

\section{Conclusions}
\label{sec:conc}
In this work, we have demonstrated a realization of emergent spacetime supersymmetry (SUSY) on a quantum computing platform by targeting the tricritical Ising (TCI) transition in the O'Brien-Fendley (OF) model. Using a variational quantum circuit, we prepared both ground and first excited states of the OF Hamiltonian. We probed the groundstate properties through supercurrent correlation functions. These measurements, performed on the IBM quantum hardware, showed consistency with the critical behavior expected from the TCI field theory, providing evidence that key signatures of emergent SUSY can be observed in near-term quantum devices. While hardware noise and limited qubit counts currently restrict access to other universal quantities, such as gap ratios, the successful preparation and verification of supersymmetric states marks significant progress in the exploration of nontrivial quantum field theory phenomena using quantum devices.

In addition to examining correlation functions, we explored the connection between Shannon entropy and the central charge in different measurement bases. Our results suggest that, despite finite-size effects, the $x$-basis Shannon entropy can serve as a useful diagnostic for extracting the central charge, yielding estimates that align with theoretical expectations for the TCI model. These findings illustrate the potential of near-term quantum devices to simulate and test emergent high-energy phenomena in condensed matter systems. As quantum hardware continues to improve, these methods lay the groundwork for future investigations of dynamical responses and the properties of highly excited states in models exhibiting SUSY that lie beyond the reach of classical computational techniques.

\begin{acknowledgments}
We thank Kartiek Agarwal and Romain Vasseur for helpful discussions. Work at Los Alamos was carried out under the auspices of the U.S. Department of Energy (DOE) National Nuclear Security Administration (NNSA) under Contract No. 89233218CNA000001. It was supported was supported by DOE Grant No. DE-SC0024641, and in part by Center of Integrated Nanotechnologies, a DOE BES user facility, in partnership with the LANL Institutional Computing Program for computational resources. We acknowledge the use of IBM Quantum services. 

\end{acknowledgments}
\appendix

\section{Correlation for small periodic system}\label{periodic}
\begin{figure}[]
\centering
\includegraphics[width=0.4\textwidth]{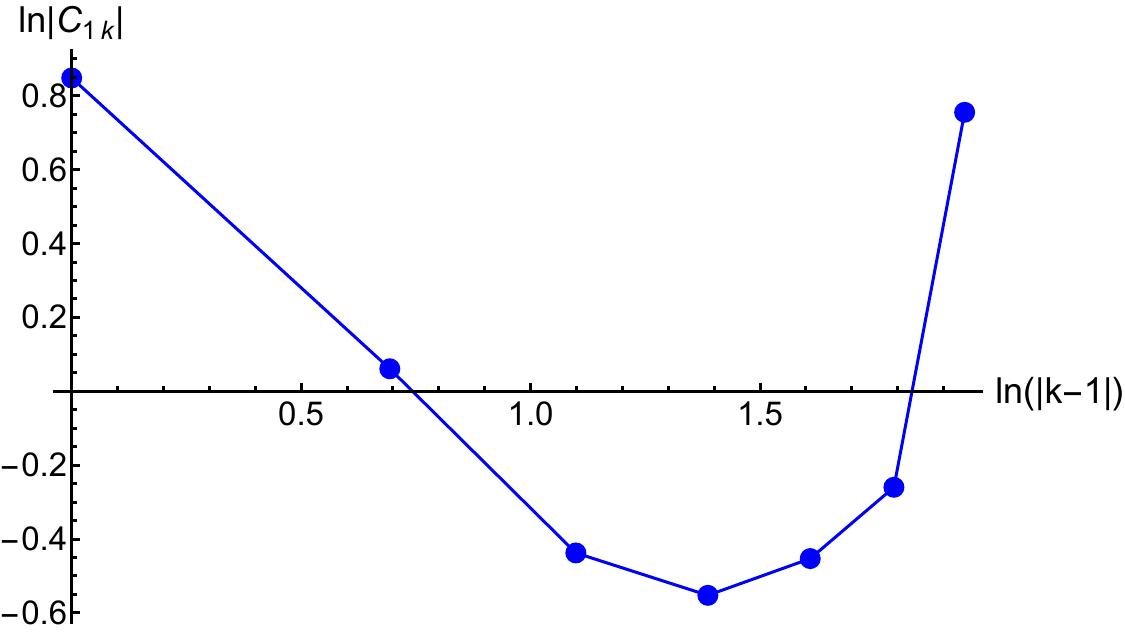}
\caption{Dependence of $C_{jk}$ on $|j-k|$ at the tricritical point for periodic boundary condition.}\label{periodicL8}
\end{figure}

In this appendix, we compute correlation $C_{jk}$ as a function of $|i-j|$ for a small system size of 8 sites with periodic boundary conditions. The correlation function is shown in Fig.~\ref{periodicL8}. The correlation decreases as a function of distance for the first half of the lattice, then it starts to increase because the two points approach each other from the other side. Therefore, more sites on the chain are affected by short-distance physics compared with open boundary conditions, which is the primary reason we carried out the demonstrations on the quantum device with open boundary conditions.

\section{Variational preparation of excited states}
Energies of excited states are valuable in the verification of SUSY.
The energy $E_\alpha$ required to excite the CFT operator with scaling dimension $\Delta_\alpha$ is proportional to the scaling dimension. Therefore, the ratios of the energy gaps are proportional to the ratios of the difference of the scaling dimensions, that is, $(E_\alpha - E_\beta)/(E_\gamma - E_\delta) = (\Delta_\alpha -\Delta_\beta)/(\Delta_\gamma - \Delta_\delta)$. The TCI CFT has six primary operators in the periodic spin chain (PBC sector) \cite{Zou:2019iwr}, $\mathbf{1}, \epsilon, \epsilon', \epsilon'', \sigma, \sigma'$.  
The spin-flip operator $\mathcal{F}$ is a symmetry of the system, and it obeys $\mathcal{F}^2 = 1$. So the Hilbert space is split into two sectors corresponding to two eigenvalues $\pm1$ of $\mathcal{F}$. The scaling dimension and parity of the primary operators are listed in Table~\ref{tab:CFToperators}. The operators are arranged in increasing values of their dimensions. 
\begin{table}[ht]
    \centering
   \begin{tabular}{|c|c|c|c|c|c|c|}
\hline
 $\phi_\alpha$ & $\mathbf{1}$ & $\sigma$ &$\epsilon$ & $\sigma'$ & $\epsilon'$& $\epsilon''$\\
\hline
$\Delta_\alpha$ & 0 &$3/40$  & $1/5$ & $7/8$ &$6/5$ & $3$ \\
\hline
$\mathcal{F}_\alpha$ &  $+$ & $-$ & $+$  & $-$& $+$ & $+$\\
\hline
\end{tabular}  
    \caption{List of CFT primaries in the PBC sector with scaling dimension and parity}
    \label{tab:CFToperators}
\end{table}

\begin{figure}[ht]
\centering
\includegraphics[width=0.4\textwidth]{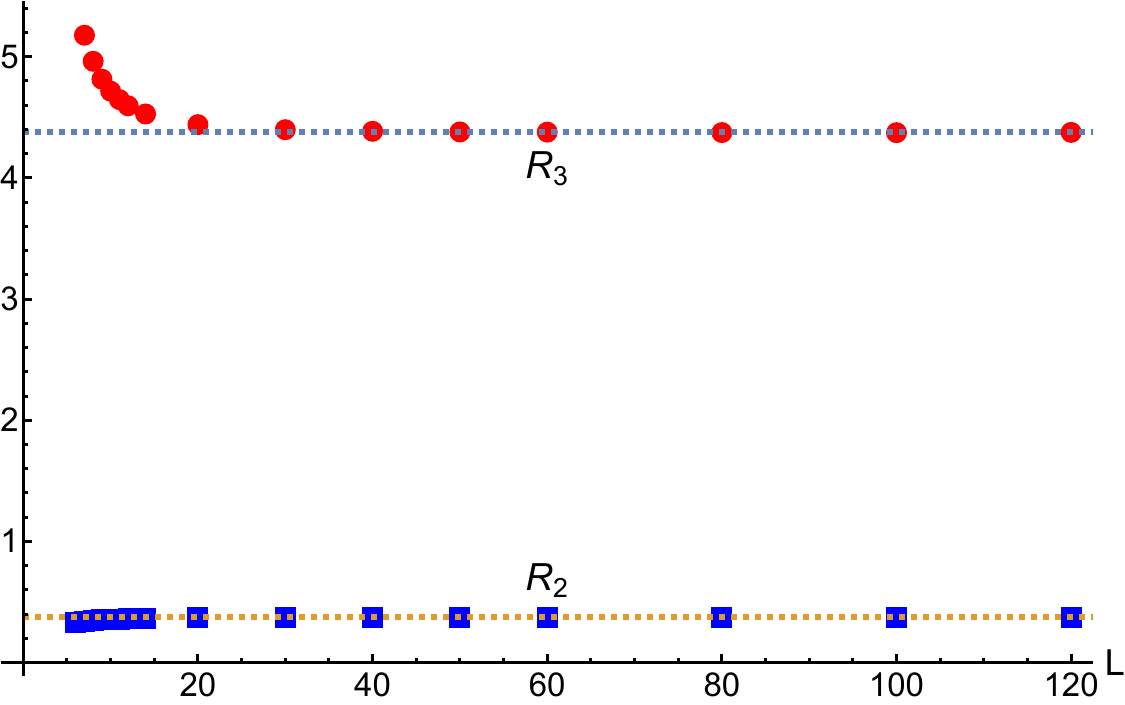}
\caption{Plot of energy ratios as a function of system size using DMRG. Theoretically predicted values are shown using dashed lines. }\label{ratio}
\end{figure}

Let us denote the ground state energy with periodic boundary conditions by $P_0^{\pm}$ and the first excited state energy by $P_1^{\pm}$. 
We can assign $P_0^+ = E_0$, $P_0^- = E_1$, $P_1^+ = E_2$, and $P_1^- = E_3$. Then we recover the energy ratios $R_2= (P_0^- - P_0^+)/(P_1^+ -P_0^+)  = 0.375$ and $R_3 = (P_1^- - P_0^+)/(P_1^+ -P_0^+)= 4.374$ in the thermodynamic limit (see Fig.~\ref{ratio}).

We aim to generate the excited states in a quantum device in order to compute these critical ratios. Variational quantum deflation (VQD) algorithm was proposed in \cite{Higgott:2018doo} to compute the excited states. The excited states can be generated in a quantum computer iteratively by minimizing the cost function
\begin{align}\label{costfunction}
    F(\lambda_k) = \langle\psi(\lambda_k)|H|\psi(\lambda_k)\rangle +  \sum_{j=0}^{k-1} \Delta_j |\langle\psi(\lambda_k)|\psi(\lambda_j)\rangle|^2\ ,
\end{align}
given that we have knowledge of $\lambda_0,\ldots, \lambda_{k-1}$. Importantly, the cost function above is, in principle, measurable on a quantum device. The expectation value can be measured directly, and the overlap requires the swap test circuit ~\cite{Higgott:2018doo}. We generate the first excited state using this VQD for the O'Brien-Fendley model with a periodic boundary condition. We obtained very accurate results for the first excited state. However, we were able to achieve an overlap of around $97\%$ and an error in energy density of around $0.03$ for the second excited state for the largest system sizes studied.
\begin{figure}[ht]
\centering
\includegraphics[width=0.44\textwidth]{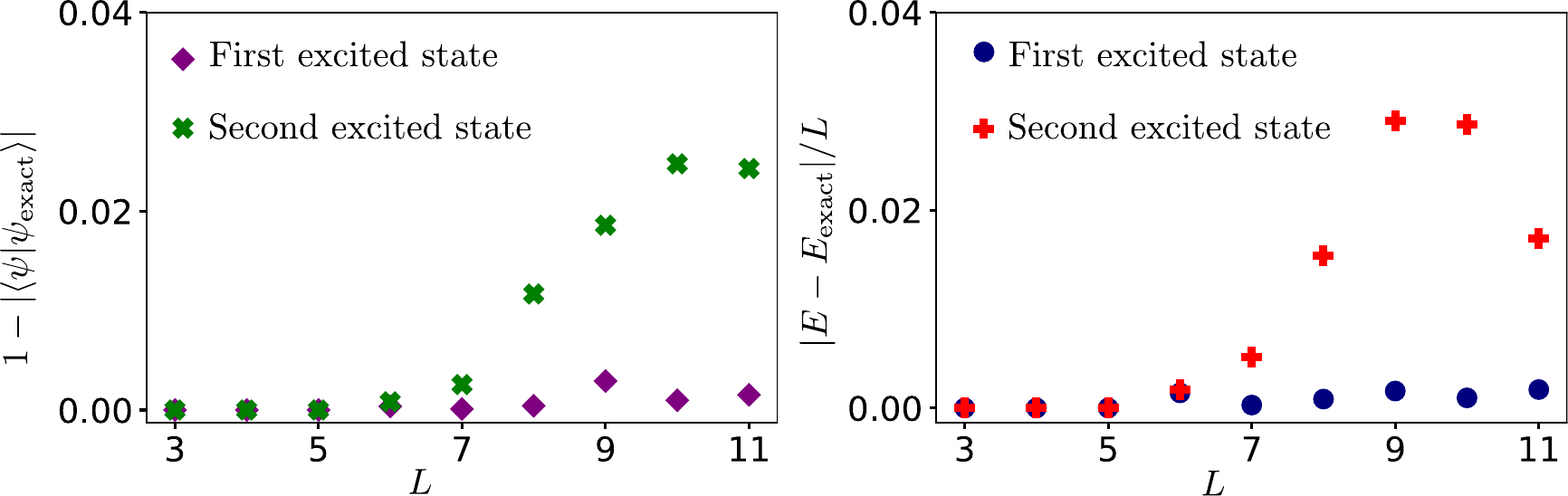}
\caption{Quality of first and second excited state generated from VQD algorithm. The left panel shows the deviation from the ideal wavefunction overlap, and the right panel shows the energy density of states. }\label{excitedstate}
\end{figure}

Universal gap ratios converge at system sizes larger than $L=10$. Although we have small errors in individual variational energies, the gap ratios are sensitive to errors in variational energies. Although our variational algorithm produces results in good agreement with exact diagonalization up to a system size of $L=6$, we find larger deviations for larger systems where the finite-size ratio begins to converge to the universal value as summarized in Table~\ref{tab:R2values}.

\begin{table}[t]
    \centering
    \begin{tabular}{|c|c|c|}
    \hline
    System size & $R_2$ Exact & $R_2$ from VQD \\
    \hline
    3 &0.236010  & 0.236010\\
    \hline
    4 &0.288045 & 0.288045\\
    \hline
    5 & 0.315029 & 0.315029\\
    \hline
    6 &0.332215  & 0.304887\\
    \hline
    7 & 0.342204 & 0.268293\\
    \hline
    8 & 0.348891 & 0.230058\\
    \hline
    9 &0.353408  & 0.191291\\
    \hline
    10 & 0.356693& 0.162929\\
    \hline
    11 &0.359155& 0.197027\\
    \hline
    \end{tabular}
    \caption{Values of $R_2$ for smaller system sizes. ``$R_2$ exact'' denotes the result computed using exact diagonalization and ``$R_2$ from VQD" denotes values computed from variational circuits.}
    \label{tab:R2values}
\end{table}

\section{Layout of the computing device and its characteristics}
\begin{figure}[t]
    \centering
    \includegraphics[width=0.48\textwidth]{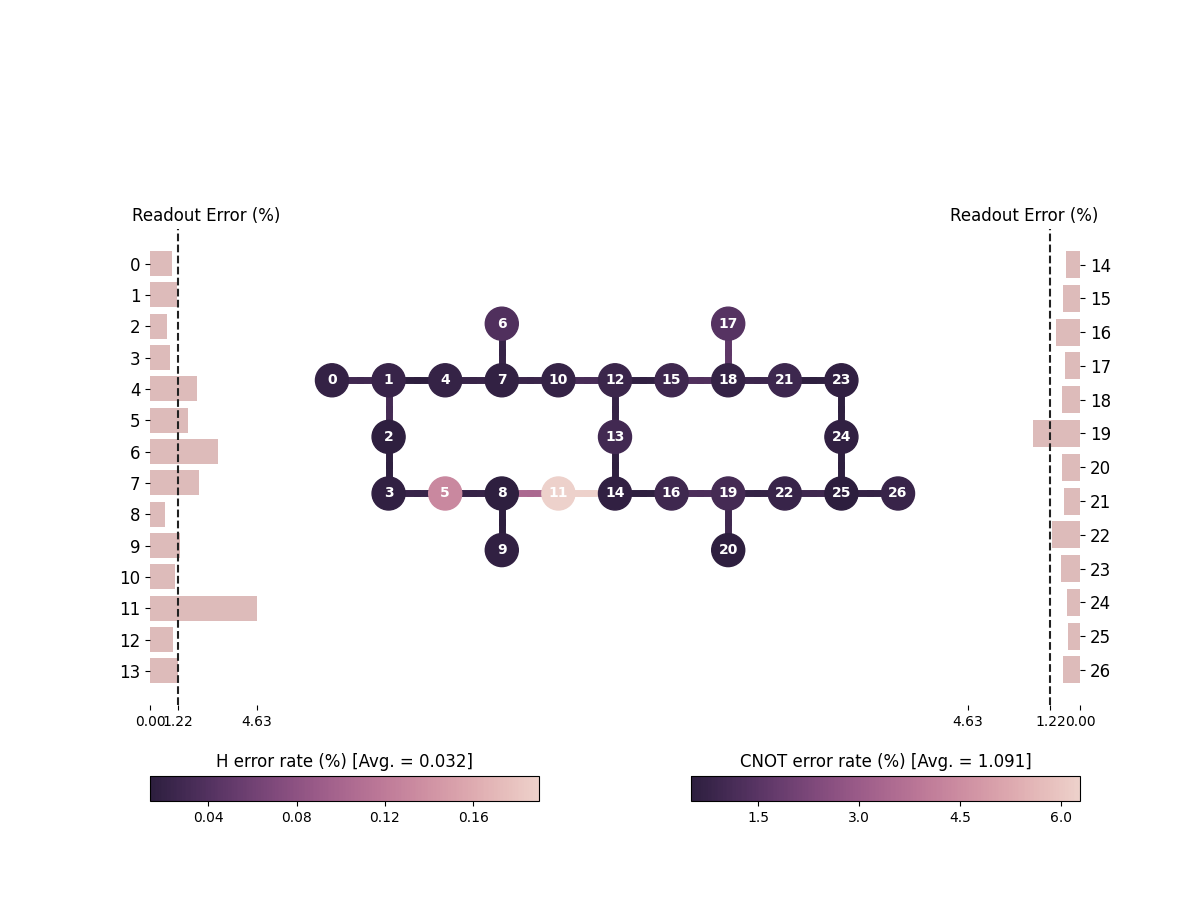}
    \caption{Qubit layout of \texttt{ibmq\_kolkata} device an its error map for qubits and couplings of \texttt{imbq\_kolkata} device.}
    \label{fig:231004}
\end{figure}
The layout of \texttt{ibmq\_kolkata} device and its error map is shown Fig.~\ref{fig:231004}. We use qubits 24, 23, 21, 18, 15, 12, 13 for our scaling law demonstrations of 7-qubit system. The readout length for all these runs was $640 ns$. $P_{01}$ ($P_{10}$) is the probability that measuring a qubit gives 0 (1) immediately after preparing it in state 1 (0). The readout error is indicated as RE, and the frequency $f$ and the anharmonicity $a$ are both in GHz.
The calibration data from October 4th, 2023, for the used qubits are given in Table~\ref{table:231004}, and the error map is shown in Fig.~\ref{fig:231004}. 
\begin{table}[h]
\begin{tabular}{ |c|c| c| c| c| c| c| c| } 
 \hline
Qubit & T1($\mu$s) & T2($\mu$s) & $f$ & $a$& RE & $P_{01}$ & $P_{10}$  \\ 
 \hline
 24&118.44 & 69.736 & 5.005 & $-0.346$ & 0.0085 & 0.011 & 0.006 \\
 \hline
 23&  158.44 & 58.638 & 5.138 & $-0.343$ &0.0055 & 0.0086 & 0.0024 \\
 \hline
 21& 25.26 & 14.382 & 5.274 & $-0.341$ & 0.0146 & 0.0152 & 0.014 \\
 \hline
 18& 138.13 & 144.38 & 5.097 & $-0.344$ & 0.0096 & 0.010 & 0.0092 \\
 \hline
 15& 65.21 & 190.24 & 5.041 & $-0.344$ & 0.0098 & 0.0136 & 0.006 \\
 \hline
 12& 108.24 & 134.36 & 4.961 & $-0.346$ & 0.0074 & 0.0094 & 0.0054 \\
 \hline
 13& 87.05 & 154.5 & 5.0178 & $-0.346$ & 0.0092 & 0.0094 & 0.009 \\
 \hline
\end{tabular}
\caption{Calibration data for the qubits used for our demonstrations. }
\label{table:231004}
\end{table}

\bibliographystyle{apsrev4-2}
\bibliography{refs}
\end{document}